\documentclass[%
 reprint,
 amsmath,amssymb,
 aps,
]{revtex4-2}
\usepackage[skip=2pt,font=scriptsize]{caption}
\usepackage{graphicx}
\usepackage{dcolumn}
\usepackage{subcaption}
\usepackage{caption}
\usepackage{amsthm}
\usepackage{bm}
\usepackage{xcolor}
\usepackage{tabularx}
\begin{document}

\title{Electric Double Layer from Phase Demixing Reinforced by Strong Coupling Electrostatics}

\author{YeongKyu Lee}
\thanks{These two authors contributed equally}
\affiliation{Physics Department, Gyeongsang National University}
\author{JunBeom Cho}
\thanks{These two authors contributed equally}
\affiliation{School of Chemical and Biological Engineering, Seoul National University}
\author{Yongkyu Lee}
\affiliation{School of Chemical and Biological Engineering, Seoul National University}

\author{Won Bo Lee}
\email{wblee@snu.ac.kr}
\affiliation{School of Chemical and Biological Engineering, Seoul National University}

\author{YongSeok Jho} 
\email{ysjho@gnu.ac.kr} 
\affiliation{Physics Department, Gyeongsang National University}

\date{\today}

\begin{abstract}
Ionic liquids (ILs) are appealing electrolytes for their favorable physicochemical properties. However, despite their longstanding use, understanding the capacitive behavior of ILs remains challenging. This is largely due to the formation of a non-conventional electric double layer (EDL) at the electrode-electrolyte interface. This study shows that the short-range Yukawa interactions, representing the large anisotropically charged ILs, demix IL to create a spontaneous surface charge separation, which is reinforced by the strongly coupled charge interaction.
The properties of the condensed layer, the onset of charge separation, and the rise of overscreening and crowding critically depend on the asymmetry of Yukawa interactions.
\end{abstract}

\maketitle
\underline{Introduction: }
Room-temperature ionic liquids (RTILs) are combinations of large organic cations and organic, inorganic anions with delocalized charges. Unlike conventional inorganic salts, these structural characteristics prevent them from forming crystalline structures at room temperature~\cite{IL1, IL2, screen}. As an electrolyte solvent at ambient conditions, RTILs offer several advantages for energy storage applications, such as low volatility, good solvent properties, high thermal stability, and environmental sustainability~\cite{adv1, adv2, application1, application2, application3, application4}, which are especially fit for future energy storage devices, supercapacitors, or ultracapacitors.
Notably, some essential features for high-technology devices (\textit{e.g.} wireless devices and electric cars), such as higher-rate energy harvesting and long-lasting supercapacitance, are related to the electric double layer (EDL) structures in RTILs~\cite{supercapacitor1, supercapacitor2, supercapacitor3, supercapacitor4, supercapacitor5}. 
While conventional point charge models predict the development of a condensation layer at very large surface charge densities, IL exhibits EDL at very low surface charge densities and even overscreening and crowding, which are very difficult to achieve without the presence of the multivalent counterions within the point charge models. 
The key differences are from the large shape of IL molecules and their delocalized charges.  
Understanding how to embody these properties in the microscopic mechanism is challenging and crucial in utilizing ILs for energy device applications~\cite{orland3, ic1, ic2, ic3}. 

Bazant, Storey, and Korneyshev (BSK)~\cite{bsk} are among the first to incorporate a short-range electrostatic correlation to the EDL. They showed that a phenomenological modification of the Poisson equation with an extra fourth-order potential gradient term, reflecting a short-range electrostatic correlation, explains the divergence of differential capacitance. They further asserted that crowding beats overscreening at higher external fields. 
Since the phenomenological description of the BSK model, efforts have been made to uncover the microscopic mechanism for unconventional EDL features.  
D\'{e}mery \textit{et al.} solved 1D lattice Coulomb gas (1D LCG) under the constant voltage ensemble. 
Their solvable one-dimensional model showed that the 1D LCG undergoes a first-order phase transition between dense and dilute phases by adjusting fugacity~\cite{1d1, 1d2}. They claimed the transition is closely related to the discontinuous jump in differential capacitance and the transition between camel- and bell-shape capacitance.

Other directions of the approaches use statistical field theory for charged many-body. Earlier mobile point charges with fixed macro-ions were treated within the framework of the statistical field theory of ion systems~\cite{coalson1, coalson2, orland1, orland2, orland3}. 
Recently, Bossa et al. adopted the free energy analysis approaches incorporating Yukawa potential to consider the symmetric steric term and successfully reproduced the fourth order potential gradient in BSK equation~\cite{bossa, coulyukawa}. They showed that the interplay between short-ranged Yukawa and long-ranged Coulomb interactions can lead to the instability of ILs near an electrode. This instability is characterized by a divergent differential capacitance, indicating a first-order transition~\cite{bossa}.

This work uncovers the microscopic origin of the unconventional EDL formation and the consequent phase transition of RTIL capacitance employing a statistical field theory framework. 
We derive free energy functionals for asymmetric ionic liquids and extracted the modified Poisson equations via saddle point approximation. Our results reveal that the phase transition is triggered by the interplay of the strong electrostatic coupling and the phase demixing due to the asymmetric short-range interaction. We validate our theoretical predictions by comparing them with molecular dynamics simulations. We further elucidate the influence of external electric fields on the phase behavior. This work paves the way for its applicability to various systems~\cite{various1, various2, various3, various4}.

\underline{Statistical Field Theory: }
We develop a lattice-based model for the electrode of surface charge density $\sigma_{e}$ at $x=0$. The number of the particle at $r$ is $n_{\pm}(r)$ and $\sum_r n_{\pm} = N_{\pm}$. The local densities of each species are $\rho_{\pm}(r)=n_{\pm}(r)/\nu$, where $\nu$ is the volume of each lattice. The local charge density is defined as $\rho_{c}(r)=\rho_{+}(r) - \rho_{-}(r)$. 

Particles interact with each other through Coulombic and Yukawa interaction:
\begin{equation}
\begin{split}
    \beta u_{+,+} &= \frac{l_{B}}{r} + a \frac{e^{-\kappa r}}{r} \\
    \beta u_{+,-} &= -\frac{l_{B}}{r} + c \frac{e^{-\kappa r}}{r} \\
    \beta u_{-,-} &= \frac{l_{B}}{r} + b \frac{e^{-\kappa r}}{r},
\end{split}\label{eq:interaction}
\end{equation}
where $l_{B}$ is the Bjerrum length, and $a, b$ and $c$ are the Yukawa interaction parameters, and $\beta=\frac{1}{k_{B}T}$. Then, the total interaction potential energy is represented by 
\begin{equation}
\begin{split}
    \beta\mathcal{U}_{tot} &= \frac{1}{2}\int{drdr'\rho_{c}(r)v_{c}(r-r')\rho_{c}(r')}\\ 
    &+ \frac{1}{2}\int{drdr'\begin{pmatrix}\rho_{+}(r)\\ \rho_{-}(r)\end{pmatrix}^{T}A_{h}v_{y}(r-r')\begin{pmatrix}\rho_{+}(r)\\ \rho_{-}(r)\end{pmatrix}},
\end{split}
\end{equation}\label{eq:total1}
where $A_{h} = \begin{pmatrix} a & c \\ c & b\end{pmatrix}$, $v_{c}(r) = l_{B}/r$, and $ v_{y}(r) = e^{-\kappa r}/r$. Similarity transformation decouples the Yukawa term with eigenstates. 

The grand canonical partition function after the Hubbard-Stratonovich transformation yields,
\begin{align}
\mathcal{Z}_{\lambda} &= \int \mathcal{D}\phi_{e}\mathcal{D}\phi_{1}\mathcal{D}\phi_{2} \exp\left(-\frac{1}{8\pi l_{B}}\int dr \left(\nabla\phi_{e}\right)^{2} \right. \nonumber \\
    &\left. -\frac{1}{8\pi\lambda_{1}}\int dr \left(\left(\nabla\phi_{1}\right)^{2} + \kappa^{2}\phi_{1}^{2}\right)\right. \nonumber \\
    &\left. -\frac{1}{8\pi\lambda_{2}}\int dr \left(\left(\nabla\phi_{2}\right)^{2} + \kappa^{2}\phi_{2}^{2}\right)\right) \label{eq:gandcanonical2} \\ 
    &\times \exp\left(\frac{1}{\nu}\int dr \ln{\left(\lambda_{s}\left(\textbf{e}_{1}+\textbf{e}_{2}\right)\right)}\right) \nonumber,
\end{align}
where $\textbf{e}_{1} = \exp\left(-i\phi_{e}-i\phi_{1}\cos\theta+i\phi_{2}\sin\theta\right)$ and $\textbf{e}_{2} = \exp\left(+i\phi_{e}-i\phi_{1}\sin\theta-i\phi_{2}\cos\theta\right)$. The free energy is,
\begin{equation}
\begin{split}
    \beta\mathcal{F} = &\int dr\left[-\frac{1}{8\pi l_{B}}\left(\nabla\psi_{e}\right)^{2} -\frac{1}{8\pi\lambda_{1}}\left\{\left(\nabla \psi_{1}\right)^{2}+\kappa^{2}\psi_{1}^{2}\right\}\right. \\
    &\left.-\frac{1}{8\pi\lambda_{2}}\left\{\left(\nabla \psi_{2}\right)^{2}+\kappa^{2}\psi_{2}^{2}\right\}\right] + \eta\left(\tilde{\textbf{e}}_{1}, \tilde{\textbf{e}}_{2}\right),
\end{split}
\end{equation}
where 
\begin{equation*}
    \eta\left(\tilde{\textbf{e}}_{1}, \tilde{\textbf{e}}_{2}\right) = -\frac{1}{\nu}\int dr \ln\left\{\lambda_{s}\left(\tilde{\textbf{e}}_{1} + \tilde{\textbf{e}}_{2}\right)\right\},
\end{equation*}
and $\tilde{\textbf{e}}_{1} = \exp\left(-\psi_{e}-\psi_{1}\cos\theta+\psi_{2}\sin\theta\right)$ and $\tilde{\textbf{e}}_{2} = \exp\left(+\psi_{e}-\psi_{1}\sin\theta-\psi_{2}\cos\theta\right)$.
The saddle-point field approximates the functional integral as, 
\begin{equation}
\begin{split}
    &\frac{1}{4\pi l_{B}}\nabla^{2}\psi_{e} = \frac{1}{\nu} \frac{\tilde{\textbf{e}}_{2}-\tilde{\textbf{e}}_{1}}{\tilde{\textbf{e}}_{1}+\tilde{\textbf{e}}_{2}} \\ \\
    &\frac{1}{4\pi \lambda_{1}}\left(\nabla^{2}-\kappa^{2}\right)\psi_{1} = \frac{1}{\nu} \frac{-\cos\theta\tilde{\textbf{e}}_{1} - \sin\theta\tilde{\textbf{e}}_{2}}{\tilde{\textbf{e}}_{1}+\tilde{\textbf{e}}_{2}} \\ \\
    &\frac{1}{4\pi \lambda_{2}}\left(\nabla^{2}-\kappa^{2}\right)\psi_{2} = \frac{1}{\nu} \frac{\sin\theta\tilde{\textbf{e}}_{1} - \cos\theta\tilde{\textbf{e}}_{2}}{\tilde{\textbf{e}}_{1}+\tilde{\textbf{e}}_{2}}.
\end{split}\label{eq:mf}
\end{equation}

When the short-range interaction is symmetric, \textit{i.e.} $a=b$, Eq.~\ref{eq:mf} yields,
\begin{equation}
\begin{split}
\frac{\nu}{4\pi l_{B}}\nabla^{2}\psi_{e} &= \tanh\left(\psi_{e} + \psi_{h}\right) \\
\frac{\nu}{2\pi(c-a)}\left(\nabla^{2}-\kappa^{2}\right)\psi_{h} &= \tanh\left(\psi_{e}+\psi_{h}\right),
\end{split}\label{eq:aeqb}
\end{equation}
where $\psi_{h} = -\psi_{2}/\sqrt{2}$. For this special case, our Eq.~\ref{eq:aeqb} is identical to the equations in previous work~\cite{bossa}. To check correspondences, we used boundary conditions~\cite{bossa, charge_oscillation}
\begin{equation}
\begin{split}
    \psi_{e}'(0) &= -\lambda s_{e} \\ \\
\begin{pmatrix} \psi_{1}'(0) - \kappa \psi_{1}(0) \\ \\ \psi_{2}'(0) - \kappa \psi_{2}(0) \end{pmatrix}
&= -4\pi A_{h}\begin{pmatrix} \sigma_{1} \\ \\ \sigma_{2}\end{pmatrix},
\end{split}\label{eq:bc}
\end{equation}
where $\lambda^{2}=4\pi l_{B}/\nu$ and $s_{e} = \lambda\sigma_{e}/e$. We set $\sigma_{1}=\sigma_{2}=0$ for the simplicity. 
The numerical solutions of these nonlinear differential equations indeed reproduce the previous results~\cite{bossa} perfectly (Fig.~\ref{fig:ref_matching}).

\begin{figure}
    \centering
    \begin{subfigure}[b]{0.45\textwidth}
         \centering
         \includegraphics[width=\textwidth]{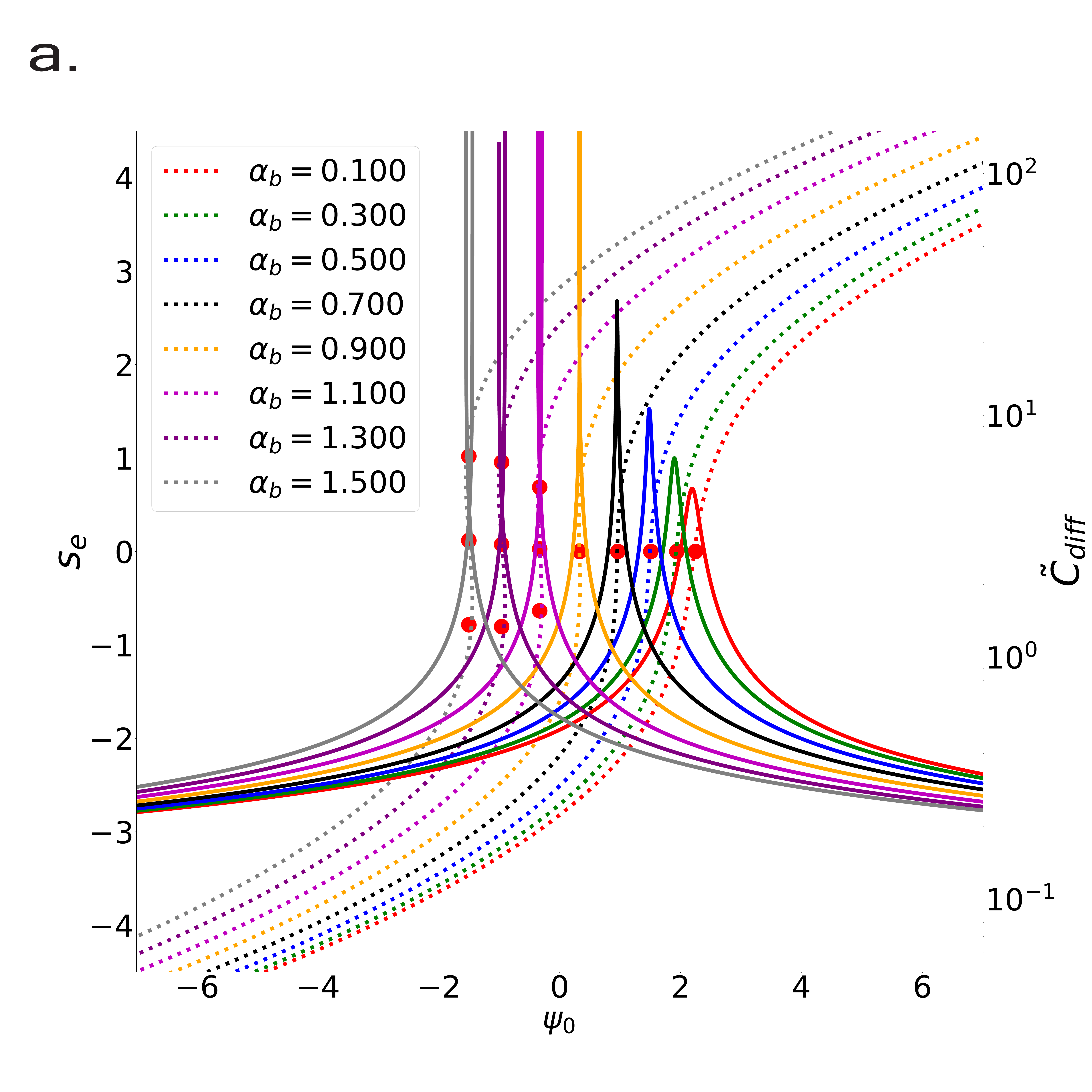}
         \label{fig:fig02a}
     \end{subfigure}
     \begin{subfigure}[b]{0.45\textwidth}
         \centering
         \includegraphics[width=\textwidth]{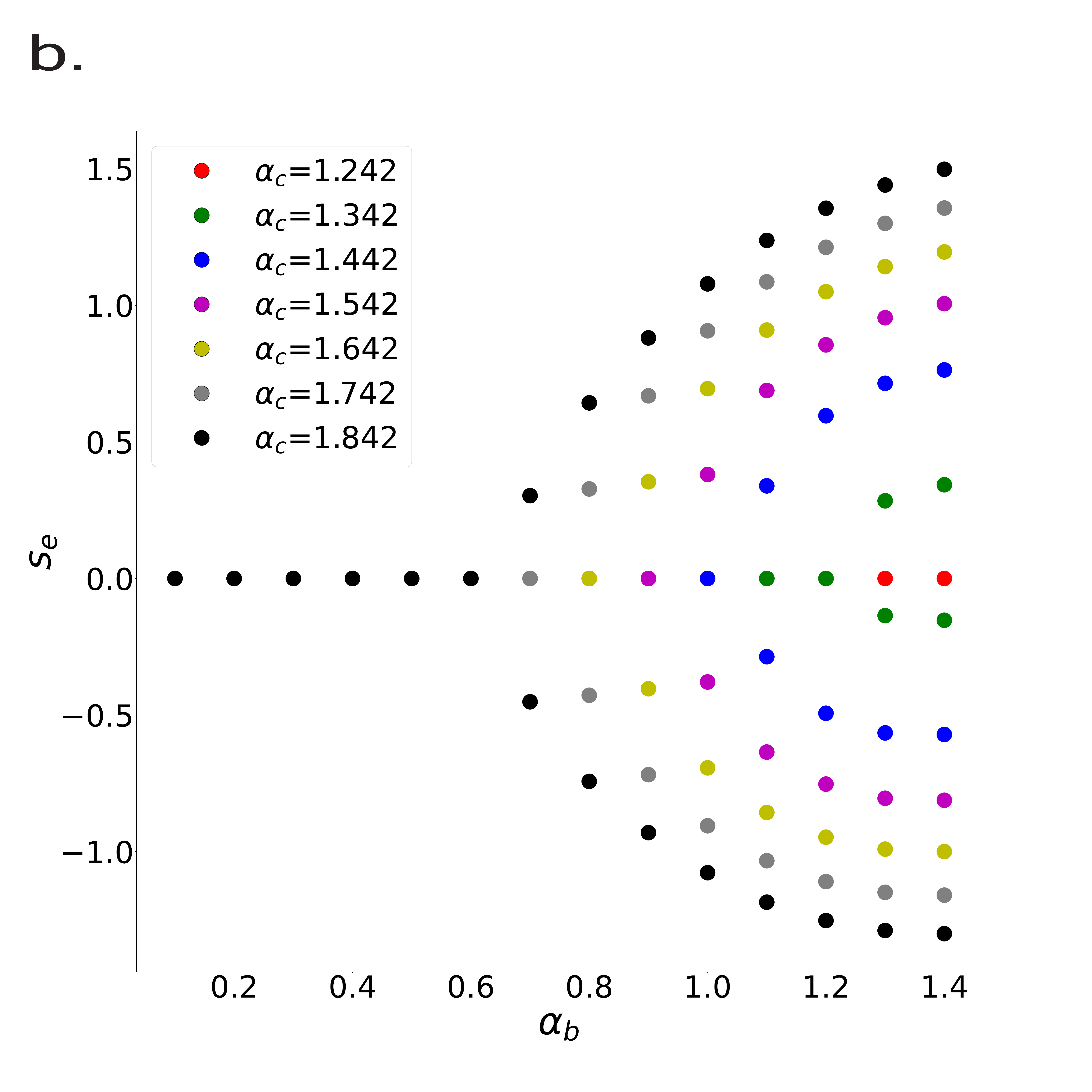}
         \label{fig:fig02b}
     \end{subfigure}
    \caption{a) Scaled surface charge density (dotted) and scaled differential capacitance (solid) as a function of surface electrostatic potential by varying $\alpha_{b}$ values. b) Scaled surface charge density as a function of $\alpha_{b}$: $\alpha_{b}$ varies from $0.1$ (most right) to $1.3$ (most left) with an increment of $0.2$. Red markers indicate $s_{e}$ at $\psi_{0}$ where $\tilde{C}_{diff}$ diverges. }
    \label{fig:fig2}
\end{figure}

To capture the behavior of the more realistic systems, we delve into asymmetric cases, $a \neq b$. We scale $b$ and $c$ relative to $a$, $b=a \times \alpha_{b}$ and $c=a \times \alpha_{c}$, respectively. 
Under the zero charge boundary condition of  $\sigma_{1}=\sigma_{2}=0$, we explore the influence of $\alpha_b$ by numerically solving Eq.~\ref{eq:mf}. Fig.~\ref{fig:fig2}.a reveals a deviation of $s_{e}\left(\psi_{0}\right)$ from the symmetric solution which transits at $\psi_{0}=0$. The emergence of hysteresis in the $s_{e}\left(\psi_{0}\right)$ curve coincides with a divergence in the differential capacitance $\tilde{C}_{diff}=\frac{1}{d \psi_{0}/d s_{e}}$, which suggests the first-order transition during EDLs formation, consistent with the previous studies~\cite{bossa, charge_oscillation}. Notably, short-range attractions induce overscreening or crowding, which is rarely observed in pure electrostatic systems. 

We next investigate the influence of the cation-anion interaction, $\alpha_c$. Fig.~\ref{fig:fig2}-b) demonstrates that increasing $\alpha_c$ shifts the onset of $\tilde{C}_{diff}$ toward lower $\alpha_{b}$. This intriguing trend can be attributed to the enhanced aggregation of like-charged ionic liquids at larger $\alpha_c$. 
Stronger cation-anion attraction elevates the energy penalty associated with mixing oppositely charged species.
We can get insight considering the effective interaction parameter $\xi = a\left(1+\alpha_b-2\alpha_c\right)$: as $\alpha_c$ increases, EDL transition becomes feasible even at lower $\alpha_{b}$.

\underline{Limiting Case: } Some limiting behaviors may provide insights. 
Setting $a=b$ allows us to recover the BSK equation. If further conditions are imposed on $c$ as $c=a=b$, $\psi_h$ is independent of the electrostatic part and exhibits exponential decay. 
When $c \gg |a-b|$, $\theta\simeq\pi/4$. Since the Yukawa interaction is only meaningful between oppositely charged ILs, we can remove one of the fields to obtain a fourth-order differential equation. 
Conversely, if $ c \ll |a-b|$, $\theta \simeq 0$. The Yukawa interactions are decoupled for each sign of charges so that one field is deduced by rewrite~\ref{eq:mf} in terms of $\psi_e+\psi_1$ and $\psi_e-\psi_2$.

\underline{Molecular Dynamics Simulation}
We perform molecular dynamics (MD) simulations employing Ye's coarse-grained model, which successfully induced the divergence of differential capacitance~\cite{benjamine}. We add a Yukawa interaction to Ye's model to compare the simulation results with our theory. 
The details of the simulation are described in SI. 

\begin{figure}
    \centering
    \begin{subfigure}[b]{0.45\textwidth}
         \centering
         \includegraphics[width=\textwidth]{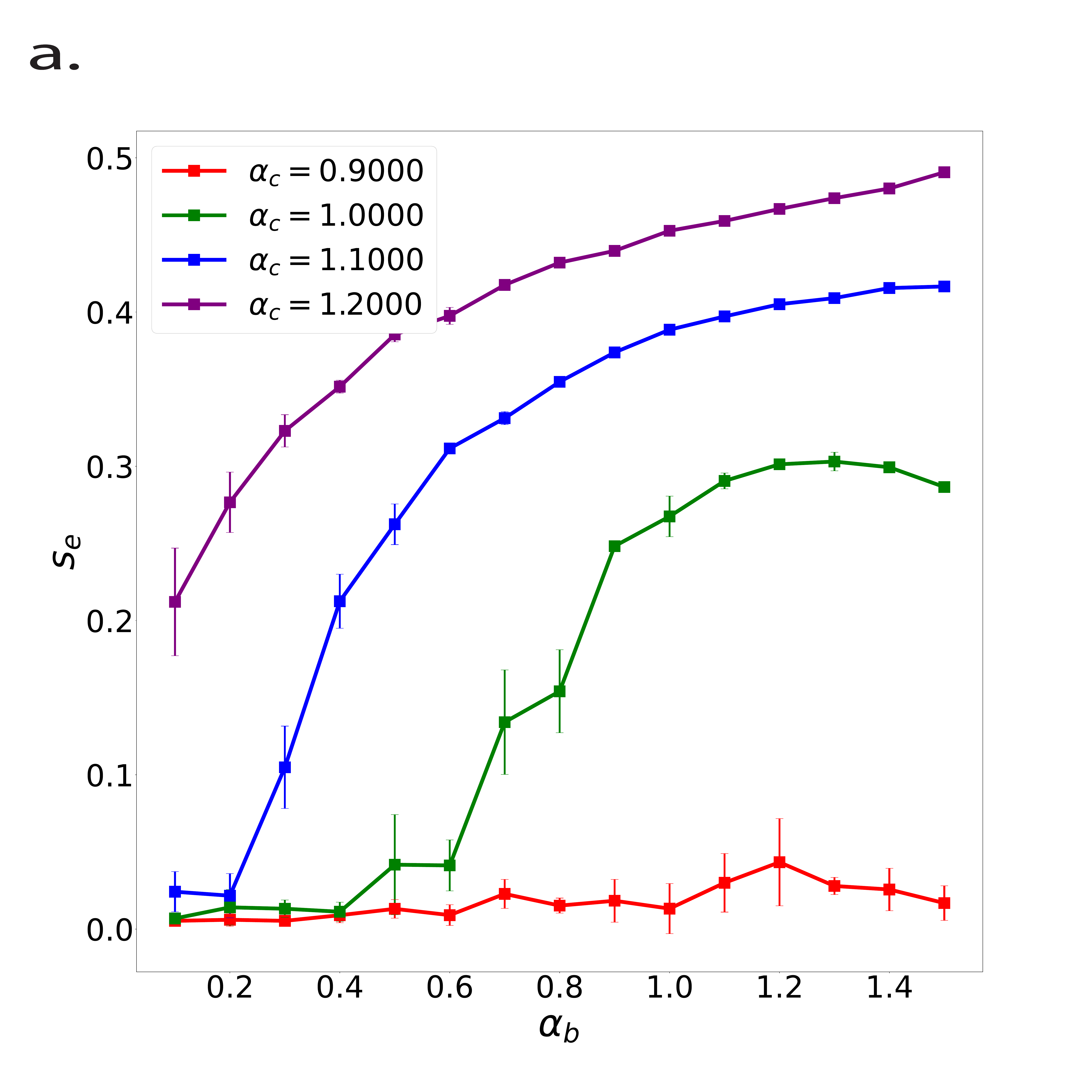}
         \label{fig:fig03a}
     \end{subfigure}
     \begin{subfigure}[b]{0.45\textwidth}
         \centering
         \includegraphics[width=\textwidth]{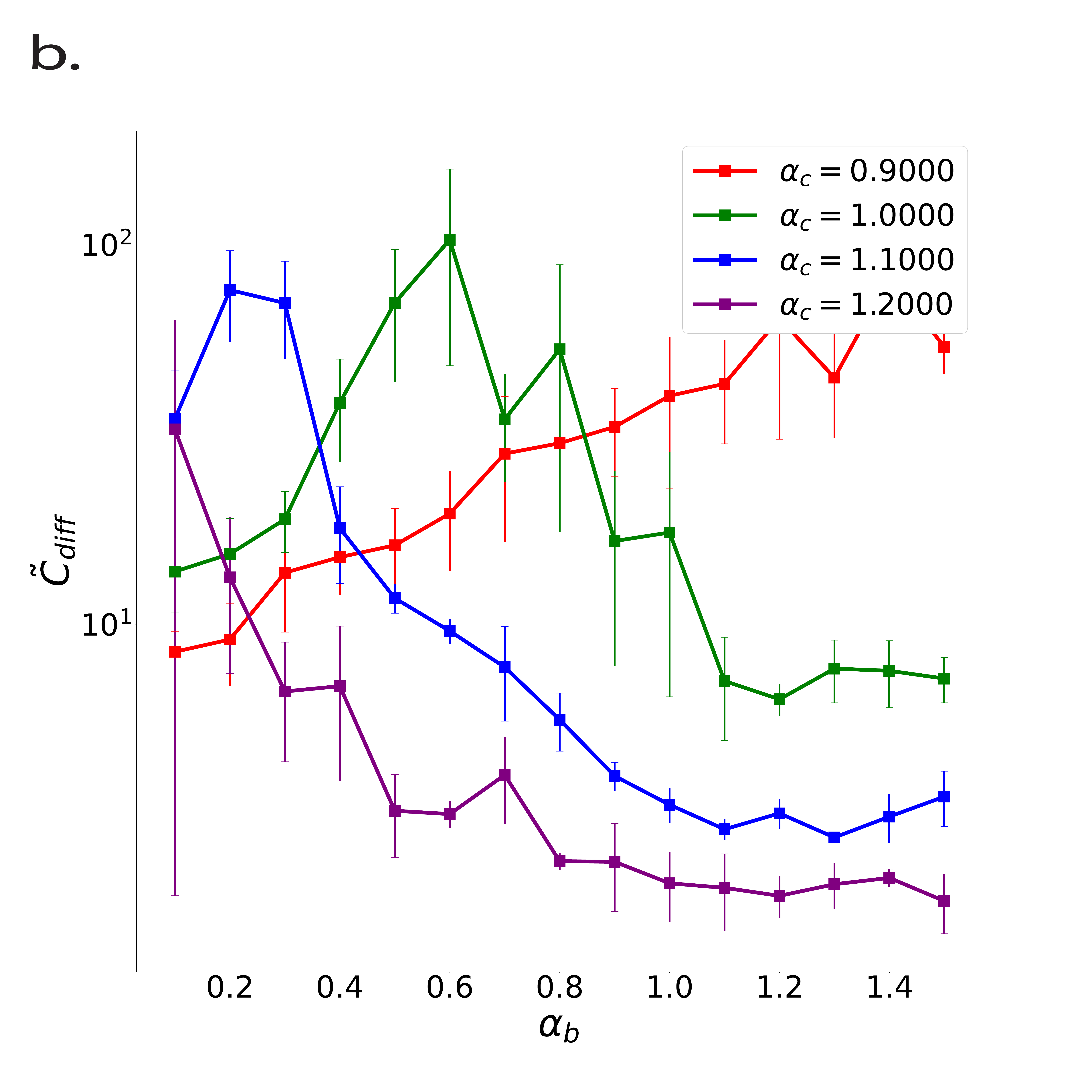}
         \label{fig:fig03b}
     \end{subfigure}
    \caption{MD results of charge separation and differential capacitance: a) No spontaneous surface charge separation (SSCS) is observed below $c=0.9$. Above $c=1.0$, SSCS occurs, and the onset of SSCS shifts toward smaller $\alpha_{b}$ consistent with the theory. b) The peak of $C_{diff}$s emerge at the onset of SSCS. Without SSCS, $C_{diff}$ shows no peak.}
    \label{fig:charge_separation}
\end{figure}

The local charge density profile, $\rho_{c}(x) = q\left(\langle n_{+}(x) \rangle - \langle n_{-}(x) \rangle \right)$, is obtained by averaging simulation configurations. $n_{+}(x)$ and $n_{-}(x)$ are the number densities of cations and anions, respectively. The densities are averaged over surface parallel directions. 
The surface charge density is calculated by, 
\begin{equation}
s_{e} = \frac{\epsilon_{r} \Delta V}{4 \pi L_{x}} - \frac{1}{L_{x}}\int_{0}^{L_{x}} x \rho_{c}(x) dx,
\end{equation}
where $\Delta V$ is the difference of constant potential between electrodes~\cite{benjamine}. 

\begin{figure}
    \centering
    \includegraphics[width=0.5\textwidth]{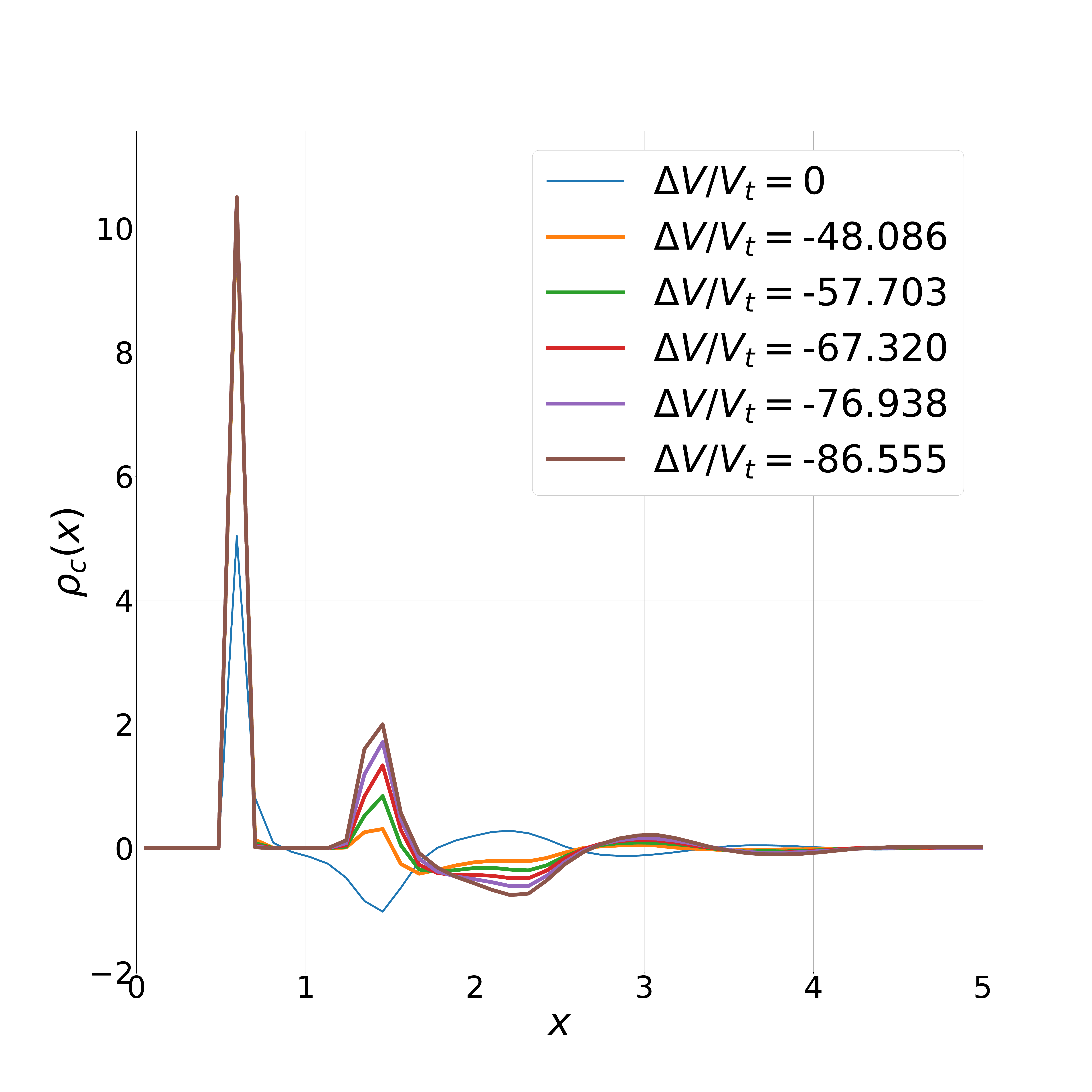}
    \caption{Charge density profiles varying external field.}
    \label{fig:ext_field_charge_density}
\end{figure}

Zero-field simulations ($\Delta V=0$) show the role of Yukawa potentials in forming EDLs without the charge coupling effects. Fig.~\ref{fig:charge_separation}.a displays that spontaneous surface charge separation (SSCS) is absent until $\alpha_{c}=0.9$. It emerges when $\alpha_{c}$ exceeds $1.0$. In addition, the onset of SSCS shifts toward lower values of $\alpha_{b}$ as $\alpha_{c}$ increases. This is analogous to our field-theoretic results shown in Fig.~\ref{fig:fig2}-b). 
Notably, despite exploring a reasonable range of Yukawa coefficients, the crowding in EDLs was absent in our zero-field simulations. The parameters in the ref~\cite{benjamine} for crowding structure lead to the void formation in our simulation, perhaps due to the Yukawa term.
Applying external potentials, the crowding appears across a wide range of the Yukawa coefficients. Fig.~\ref{fig:ext_field_charge_density} shows the crowding emerges above a certain external potential threshold, while overscreening dominates below it. 

Previous works, such as ref~\cite{bsk, kornyshev_md}, have identified the external field as a crucial factor governing the transition between overscreening and crowding. Interestingly, our model predicts overscreening, crowding, and bulk phase demixing despite low external fields. The short-range Yukawa interaction plays a critical role in spontaneous surface charge separation and multiple-layer formation inside EDL.

Purely electrostatic systems do not form an electric double layer unless the surface charge density exceeds $1e/nm^2$ or multivalent counter-ions exist. 
However, ionic liquids exhibit EDL formation even at very low voltages, suggesting an additional driving force beyond simple charge coupling. 
The multipolar interactions and hardcore repulsion arising from the large, delocalized charges within the ionic liquid induce an alternative way of EDL formation.  

Our model captures these non-electrostatic interactions through Yukawa interactions. In the limit that molecules carry zero charges, the effective short-range interaction may be described by $\chi = a\left(1+\alpha_b-2\alpha_c\right)$. But, in the opposite limit, for example, $\kappa=\infty$, $\phi_1$, and $\phi_2$ are zero, the coupling parameter $\Xi = q^2 \frac{l_B}{\mu}$ governs counter-ion condensation. In between, both parameters come into play, influencing the EDL formation. 

A negative value of $\chi$ leads to spontaneous surface charge separations.
At the beginning of the simulation, two charges form local domains at both electrodes, which may be metastable due to large line tensions~\cite{benjamine}. 
Given enough time, the metastable state transits to a stable state, in which a single charge species is abundant in each electrode~\cite{benjamine, atkin}. The condensed layer easily overscreens the surface charges, which is extremely difficult to achieve for pure electrostatic cases. 

In our case, because the Yukawa coefficients modify the electrostatic interactions also, $\chi$ cannot fully explain the surface phase demixing. For instance, increasing $a$ or $b$ has a distinct effect compared to decreasing $c$, even though both yield the same $\chi$. This is because they also adjust the Coulombic interactions (in other words, the total interaction differs for both cases.). Smaller $\alpha_b$ makes large anionic domain formation difficult, while larger $\alpha_c$ enhances it. 
If $\alpha_c$ is unreasonably large enough to ignore the Coulomb interaction, it provokes a bulk phase transition of the binary demixing. 

Within the parameter range explored in this study, an external field can induce the development of multiple layers of charges, also known as crowding, at the surface (see Fig.~\ref{fig:ext_field_charge_density}).
While a large surface charge density is required to trigger crowding~\cite{Kornyshev, nmat}, the onset voltage varies depending on the Yukawa interactions. Stronger short-range repulsion between opposite charges or attraction between like charges reduces the onset voltage~Fig.~S2. These findings highlight the interplay between the multipolar nature of the ionic liquid (which translates into the strength of the short-range interaction) and the external electric field in determining the overscreening and crowding inside EDL.

\begin{figure}
    \centering
    \includegraphics[width=0.5\textwidth]{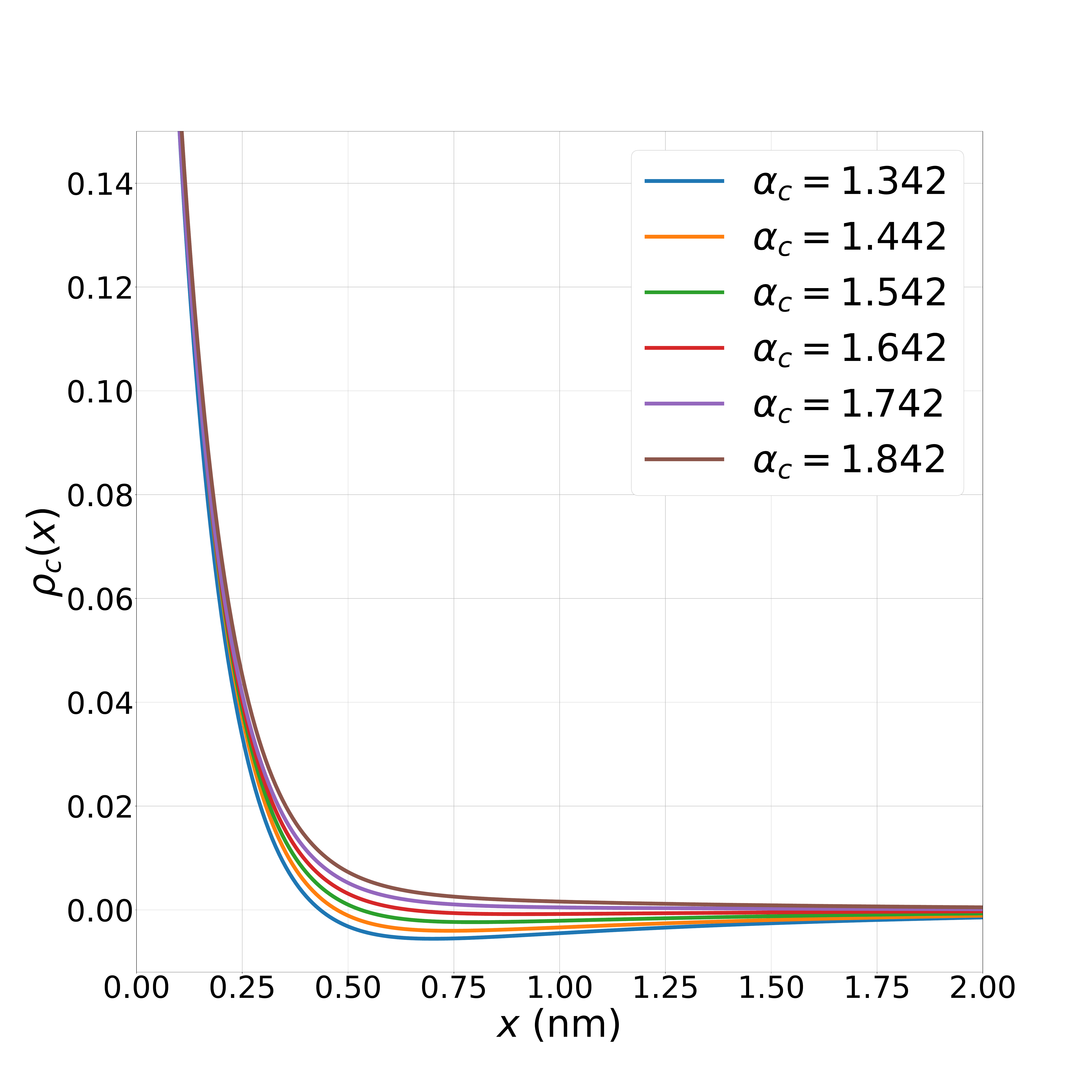}
    \caption{Local charge density profiles at the onsets of the $\tilde{C}_{diff}$ divergence}
    \label{fig:local_charge_density}
\end{figure}

The internal structure of EDL at the divergence of $\tilde{C}_{diff}$ is controversial~\cite{bsk}. At the onset of divergence, our model reveals that the counter-ion arrangement is not determined by a single EDL structure, such as underscreening, overscreening, or crowding. To better understand, we investigate the local charge density profiles at the onset of $\tilde{C}_{diff}$ divergence using our model. 
The average number of ions is  
\begin{equation}
    N_{\pm} = \lambda_{\pm}\frac{\partial \ln{\mathcal{Z}_{\lambda}}}{\partial \lambda_{\pm}}.
\end{equation}
The local charge density profile is obtained from the loop expansion adopted the zeroth order,
\begin{equation}
    \rho_{c}(r) = \sum_{i=\pm} \frac{N_{i}}{V} = \left\langle \frac{\tilde{e}_{1} - \tilde{e}_{2}}{\tilde{e}_{1} + \tilde{e}_{2}} \right\rangle = \frac{\bar{e}_{1} - \bar{e}_{2}}{\bar{e}_{1} + \bar{e}_{2}},
\end{equation}
where the potential in $\bar{e}_{1,2}$ is used as saddle-point solutions of $\psi_{e}, \psi_{1}$, and $\psi_{2}$. From the local charge density profiles in Fig.~\ref{fig:local_charge_density}, we find that condensed counterions underscreen the surface charge at $c \geq 1.742$. They start to
overscreen the surface charge at $1.342 \leq c \leq 1.642$. It is plausible that large repulsion between different species may induce spontaneous surface charge separation even at low surface charge densities and, hence, below the overscreening condition. 

This study delves into the intricate mechanisms of EDL formation at the surface of ionic liquids. It unveils the delicate interplay between two key interactions, the long-range Coulomb potential and the short-range Yukawa potential. The Yukawa potential arises from IL's large delocalized charges. 

The Yukawa potential induces spontaneous surface charge separations (or charge condensation), and the Coulomb force enhances the condensation to develop further overscreening or crowding by reinforcing the strong coupling of counter-ions at the surface. 
At the onset of the divergence of differential capacitance, SSCS occurs. At this critical point, the external field may increase the condensed charges, but the coions would compensate them immediately. 

No unique internal structure of EDLs arises at the onset of divergence of $\tilde{C}_{diff}$. The strength of short-range ion-ion correlations and surface charge density orchestrate the internal structure of EDLs. Considering these ionic liquids' properties, we can apply these results to real-world problems such as supercapacitors and solid-state electrolyte design. 

\renewcommand\thesection{}
\section{Author Contribution}
This section is intentionally left empty.

\section{Data Availability}
The authors will provide data on reasonable request. 

\section{Acknowledgement}
This research was supported by the National Research Foundation of Korea(NRF) grant funded by the Korea government(MSIT) 2018M3D1A1058633, and 2021R1A2C1014562.

\bibliography{mybib}
\clearpage
\begin{widetext}
\newpage
\renewcommand\thefigure{S\arabic{figure}}
\section{Supporting Information}
\setcounter{figure}{0}
\renewcommand\theequation{S\arabic{equation}}
\setcounter{equation}{0}

\subsection{Reduced Lennard-Jones Units}
\begin{table}[ht]
\begin{tabularx}{\linewidth}{|X|X||X|X|}
\hline
Distance & $r^*=r/\sigma$ & Charge density & $\rho^*=\rho_{q}\sigma^3/\sqrt{4\pi\epsilon_{0}\sigma\epsilon}$ \\
Energy & $U^*=U/\epsilon$ & Surface charge density & $\sigma_{e}^*=\sigma_{e}\sigma^2/\sqrt{4\pi\epsilon_{0}\sigma\epsilon}$ \\
Number density & $\rho^*=\rho \sigma^3$ & Electric dipole & $M^*=M/\sigma\sqrt{4\pi\epsilon_0\sigma\epsilon}$ \\
Temperature & $T^*=k_{B}T/\epsilon$ & Electric field & $E^*=E\sigma\sqrt{4\pi\epsilon_0\sigma\epsilon}/\epsilon$ \\
Time & $t^*=t\sqrt{\epsilon/m\sigma^2}$ & Voltage & $V^*=V\sqrt{4\pi\epsilon_0\sigma\epsilon}/\epsilon$ \\
Charge & $q^*=q/\sqrt{4\pi\epsilon_{0}\sigma\epsilon}$ & Differential capacitance & $C_d^*=C_d\sigma/4\pi\epsilon_0$ \\
\hline
\end{tabularx}
$\epsilon$ is an energy scale.
\end{table}

\subsection{Pair Potentials of Molecular Dynamics Simulation}

This study uses four potentials: the Coulomb, Yukawa, WCA (Weeks-Chandler-Andersen), and Gaussian potentials. Particle-particle particle-mesh (PPPM) is employed to calculate Coulomb potential. The interactions apply between all charges, regardless of whether they are real or image charges. The potential has the form
\begin{equation}
    U_{Coul}(r_{ij}^{*}) = \frac{q_{i}^{*}q_{j}^{*}}{\epsilon_{r}r_{ij}^{*}},
    \label{eq:coul}
\end{equation}
where $q_{i}^{*}$ and $q_{j}^{*}$ are the reduced charges of ions $i$ and $j$, respectively, $\epsilon_{r}$ is the relative permittivity, and $r_{ij}^{*}$ is distance between $i-j$ pair in a reduced unit.

The Yukawa potential is
\begin{equation}
    U_{Y}(r_{ij}^{*}) =
    \begin{cases}
        \frac{C_{ij}}{r_{ij}^{*}}\left(e^{-r_{ij}^{*}/l_{c}} - e^{-r_{c}^{*}/l_{c}}\right), & \,\,\,\,r_{ij}^{*} \leq r_{c}^{*} \\
        0, & \,\,\,\,r_{ij}^{*} > r_{c}^{*}
    \end{cases}
    \label{eq:yukawa}
\end{equation}
where $C_{ij}$ represent the Yukawa coefficients between species of $i$th particle and $j$th particle. It is $a$ for anion-anion, $b$ for cation-anion, and $c$ for cation-cation. During the simulations, $\alpha$, is fixed to 0.75 and $b$ and $c$ are varied as $0.1a\leq b \leq 1.4a$ and $0.9a\leq c \leq1.2a$. They corresponds to $\alpha_{b}$ and $\alpha_{c}$ in theoretical formula. $l_{c}$ is the inverse decay length of the Yukawa potential set to be the ion size in this work. The cutoff distance is set to $r_{c}=L_{x}/2$.

The excluded volume interactions between real ions are accounted for by a mixture of WCA and Gaussian potentials. 
\begin{equation}
    U_{ex}(r_{ij}^{*}) = U_{WCA}(r_{ij}^{*}) + U_{Gauss}(r_{ij}^{*}).
\end{equation}
The WCA potential is 
\begin{equation}
    U_{WCA}(r_{ij}^{*}) =
    \begin{cases}
        4\epsilon_{ij}\left[\left(\frac{\sigma_{ij}}{r_{ij}^{*}}\right)^{12} - \left(\frac{\sigma_{ij}}{r_{ij}^{*}}\right)^{6}\right] + \epsilon_{ij}, & r_{ij}^{*} \leq 2^{1/6} \\
        0, & r_{ij}^{*} > 2^{1/6}
    \end{cases}
    \label{eq:wca}
\end{equation}

where $\epsilon_{ij}$ and $\sigma_{ij}$ are the scaled dispersion energy and size of ions, respectively, which are set to $\epsilon_{ij}=0.01$ and $\sigma_{ij}=1$. The shifted Gaussian potential is
\begin{equation}
    U_{Gauss}(r_{ij}^{*}) =
    \begin{cases}
        A_{ij}\left(e^{-B_{ij}r_{ij}^{*2}}-e^{-B_{ij}}\right), & \,\,\,\,r_{ij}^{*} \leq 1 \\
        0, & \,\,\,\,r_{ij}^{*} > 1,
    \end{cases}
    \label{eq:gauss}
\end{equation}
where $A_{ij}$ is interaction energy and $B_{ij}$ is interaction range which are set to $A_{ij}=1000$ and $B_{ij}=7$ following Ye’s work. 

These purely repulsive soft potentials accurately capture the excluded volume interactions of RTILs, preventing an unrealistic liquid-to-solid phase transition that can occur in densely packed simulations with hard spheres. 

The WCA potential models the non-electrostatic interaction between ions and the electrode. It differs from the excluded volume interaction with cutoff distance$r_{c}=2^{-5/6}$ and parameters $\epsilon_{ij}=100$ and $\sigma_{ij}=0.5$. The distance between an ion and the electrode is simply measured along the $x$-axis.

\subsection{Charge Density Profiles under the External Fields from the Molecular Dynamics Simulation}
We calculate charge density profiles from the molecular dynamics simulation at the same $c$ and $\alpha_{aa}$ value. In this parameter range, EDLs is overscreened even in the absence of the external field. It turns to crowding applying external fields.
As shown in main text, crowding occurs at low external field as the correlation between co-ions is stronger. 

\begin{figure}[htbp]
    \centering
    \includegraphics[width=\textwidth]{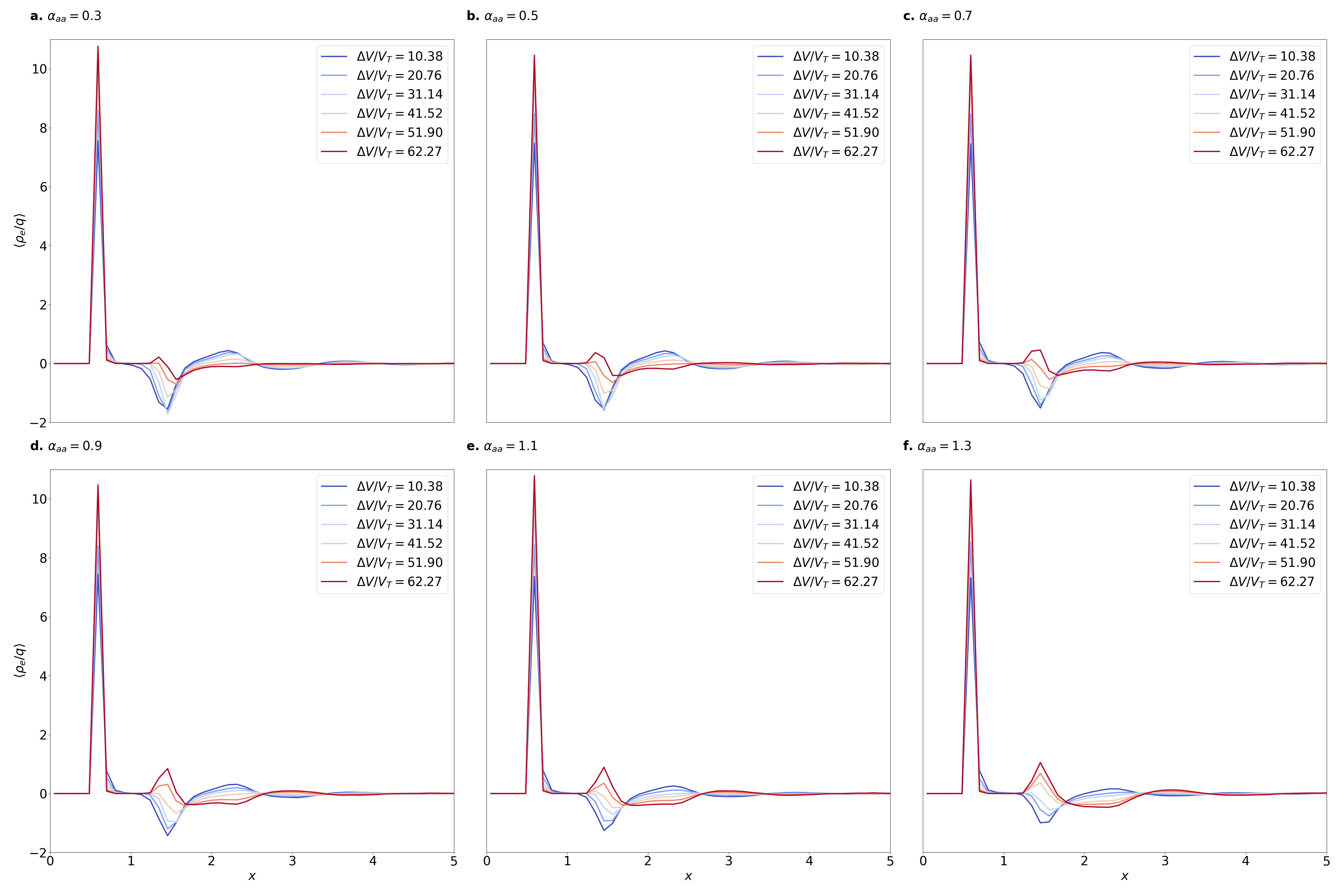}
    \caption{Charge density profile calculated from simulations applying potential differences with respect to $\alpha_{aa}$}
    \label{fig:crowding_threshold}
\end{figure}

\subsection{Reproduction of reference data}
In Fig.~\ref{fig:ref_matching}, we compare our results with previous report~\cite{bossa, coulyukawa}. We find that previous results correspond to a specific case of ours. 
\begin{figure}[htbp]
    \centering
    \includegraphics[width=\textwidth]{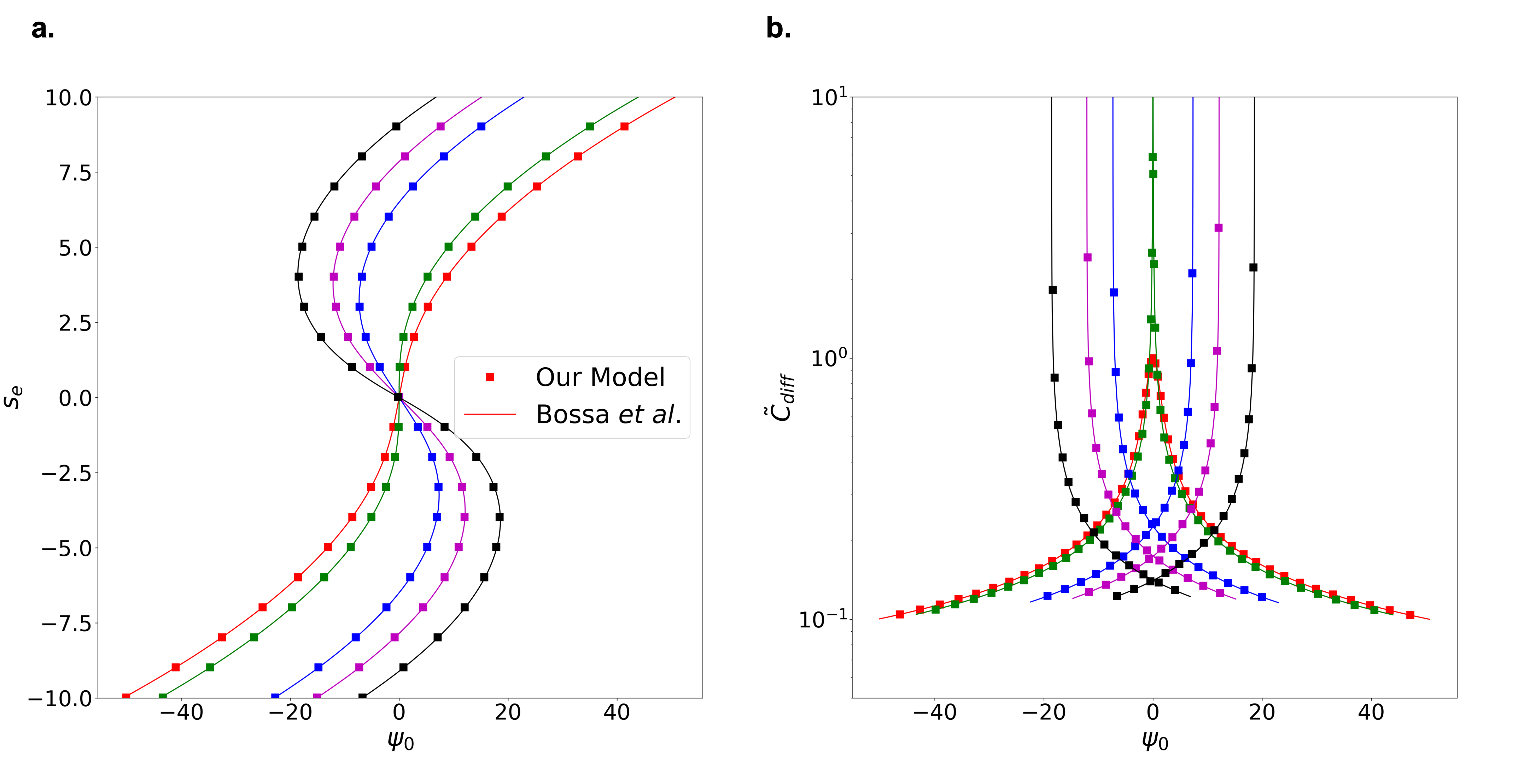}
    \caption{\textbf{a.} Scaled surface charge ($s_{e}$) and \textbf{b.} differential capacitance ($\tilde{C}_{diff}$) as a function of surface electrostatic potential $\psi_{0}$, where $\psi_{0}=\psi_{e}(0)$; The results completely matches with the work of Bossa \textit{et al.} in the case of $a=b$. We use $l_{B}$ = 8 nm, $a=b$ = $-2$ nm. $c=-2$ nm (red), $c=1.342$ nm (green), $c=10.892$ nm (blue), $c=14.075$ nm (purple), and $c=17.258$ nm (black).}
    \label{fig:ref_matching}
\end{figure}

\subsection{Crowding Structures}
\begin{figure}[htbp]
    \centering
    \includegraphics[scale=0.1]{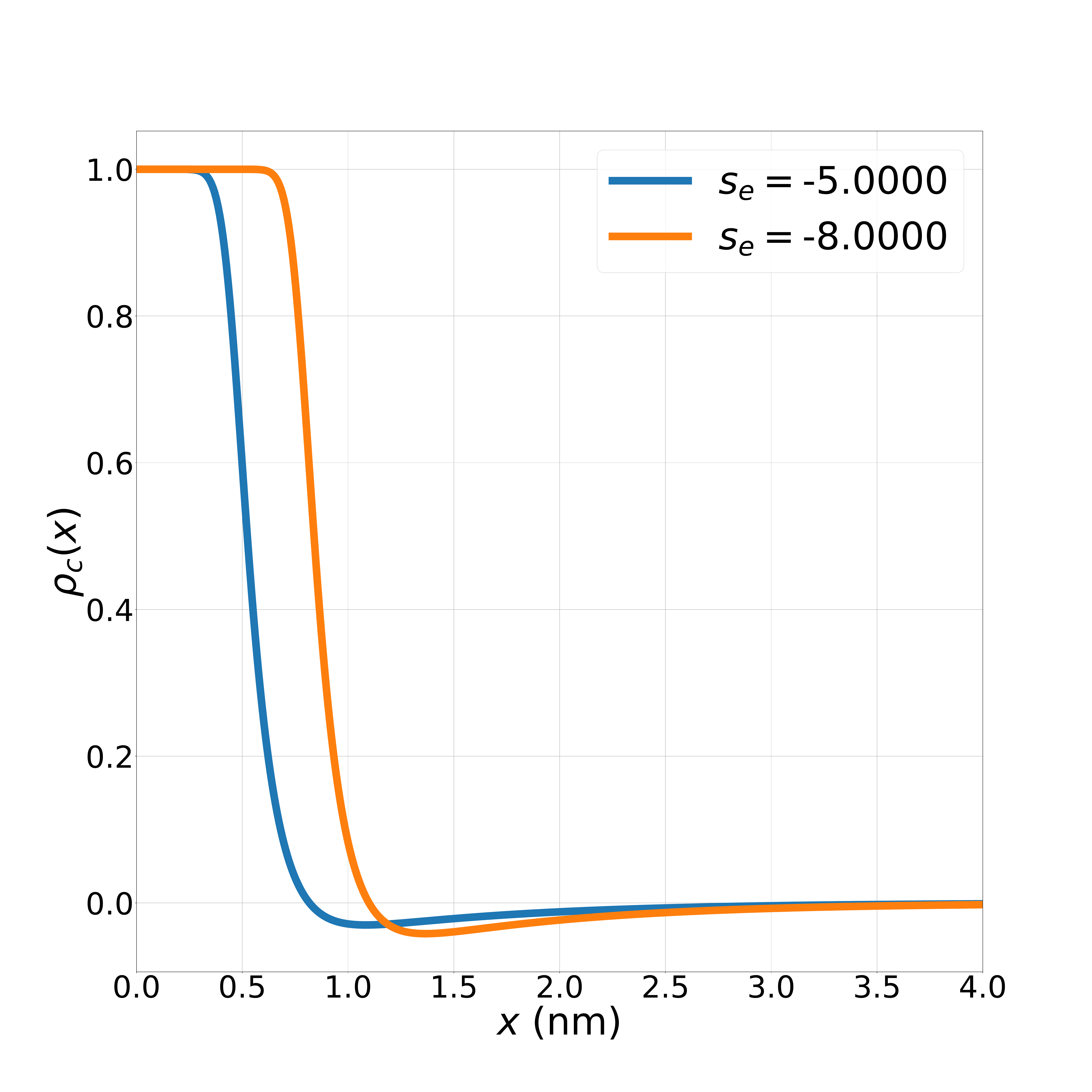}
    \caption{Local charge density profiles for the strong cation-anion short-range interaction}
    \label{fig:local_charge_density_overscreening}
\end{figure}
As the BSK model, we find crowding structure if surface charge density is high~\cite{bsk}.

\subsection{Statistical Field Theory}
We develop a lattice based model for electrode of surface charge density $\sigma_{e}$ at $x=0$. The number of the particle at $r$ is $n_{\pm}(r)$ and $\sum_r n_{\pm} = N_{\pm}$. The local densities of each species are $\rho_{\pm}(r)=n_{\pm}(r)/\nu$, where $\nu$ is the volume of each lattice. The local charge density is defined as $\rho_{c}(r)=\rho_{+}(r) - \rho_{-}(r)$. 

Particles interact each other through Coulombic and Yukawa interaction:
\begin{equation}
\begin{split}
    \beta u_{+,+} &= \frac{l_{B}}{r} + a \frac{e^{-\kappa r}}{r} \\
    \beta u_{+,-} &= -\frac{l_{B}}{r} + c \frac{e^{-\kappa r}}{r} \\
    \beta u_{-,-} &= \frac{l_{B}}{r} + b \frac{e^{-\kappa r}}{r},
\end{split}\label{eq:interaction-s}
\end{equation}
where $l_{B}$ is the Bjerrum length, and $a, b$ and $c$ are the Yukawa interaction parameters, and $\beta=k_{B}T$. Then, the total interaction potential energy is represented by 
\begin{equation}
\begin{split}
    \beta\mathcal{U}_{tot} &= \frac{1}{2}\int{drdr'\rho_{c}(r)v_{c}(r-r')\rho_{c}(r')}\\ 
    &+ \frac{1}{2}\int{drdr'\begin{pmatrix}\rho_{+}(r)\\ \rho_{-}(r)\end{pmatrix}^{T}A_{h}v_{y}(r-r')\begin{pmatrix}\rho_{+}(r)\\ \rho_{-}(r)\end{pmatrix}},
\end{split}
\end{equation}\label{eq:total1-s}
where $A_{h} = \begin{pmatrix} a & c \\ c & b\end{pmatrix}$, $v_{c}(r) = l_{B}/r$, and $ v_{y}(r) = e^{-\kappa r}/r$. Similarity transformation decouples the Yukawa term with eigenstates,
\begin{equation}
    \begin{split}
        \beta\mathcal{U}_{tot} &= \frac{1}{2}\int{drdr'\rho_{c}(r)v_{c}(r-r')\rho_{c}(r')} \\
                          &+ \frac{1}{2}\int{drdr'\rho_{1}(r)v_{y,1}(r-r')\rho_{1}(r')} \\
                          &+ \frac{1}{2}\int{drdr'\rho_{2}(r)v_{y,2}(r-r')\rho_{2}(r')},
    \end{split}
\end{equation}\label{eq:total2-s}
where $v_{y,i}(r) = \lambda_{i}v_{y}(r)$ for $i=1,2$. The $\lambda_{i}$ is
\begin{equation*}
    \begin{split}
        \lambda_{1}&=a\cos^{2}\theta+2c\sin\theta\cos\theta+b\sin^{2}\theta \\ 
        \lambda_{2}&=a\sin^{2}\theta-2c\sin\theta\cos\theta+b\cos^{2}\theta,
    \end{split}
\end{equation*}
and $\rho_{i}(r)$ is
\begin{equation*}
    \begin{split}
        \rho_{1}(r)&=\rho_{+}(r)\cos\theta+\rho_{-}(r)\sin\theta \\
        \rho_{2}(r)&=-\rho_{+}(r)\sin\theta+\rho_{-}(r)\cos\theta.
    \end{split}
\end{equation*}
Here, $\theta$ satisfies, 
\begin{equation*}
    \tan 2\theta = \frac{2c}{a-b}.
\end{equation*}

The canonical partition function of the system is,
\begin{equation}
    \begin{split}
    \mathcal{Z}_{c} &= \sum\limits_{\{n_{\pm}(r)\}}\delta\left(\sum_{r}n_{+}(r)-N_{+}\right)\delta\left(\sum_{r}n_{-}(r)-N_{-}\right)e^{-\mathcal{U}_{tot}} \\
    &=\sum_{\{n_{\pm}(r)\}}
    \delta\left(\sum_{r}n_{+}(r)-N_{+}\right)
    \delta\left(\sum_{r}n_{-}(r)-N_{-}\right) \\
    &\times\exp\left(-\frac{1}{2}\int{drdr'\rho_{c}(r)v_{c}(r-r')\rho_{c}(r')}\right. \\
    &\left.\quad\quad\quad\,\,-\frac{1}{2}\int{drdr'\rho_{1}(r)v_{y,1}(r-r')\rho_{2}(r')}\right. \\
    &\left.\quad\quad\quad\,\,-\frac{1}{2}\int{drdr'\rho_{1}(r)v_{y,2}(r-r')\rho_{2}(r')}\right).
    \end{split}
    \label{eq:canonical-s}
\end{equation}

It is more practical to deal with the grand canonical ensemble, 
\begin{equation*}
\begin{split}
\mathcal{Z_{\lambda}}=\sum_{N_{+}=0}^{\infty}\sum_{N_{-}=0}^{\infty}
\frac{\lambda_{+}^{N_{+}}}{N_{+}!}\frac{\lambda_{-}^{N_{-}}}{N_{-}!}\mathcal{Z}_{c}.
\end{split}
\end{equation*}
where fugacities $\lambda_{\pm}$ are function of the chemical potential $\mu_{\pm}$, $\lambda_{\pm} = e^{\mu_{\pm}}$.

The Hubbard-Stratonovich transformations setting $\lambda_{\pm}=\lambda_{s}$ for the charge neutrality condition yields,
\begin{align}
\mathcal{Z}_{\lambda} &= \int \mathcal{D}\phi_{e}\mathcal{D}\phi_{1}\mathcal{D}\phi_{2} \exp\left(-\frac{1}{8\pi l_{B}}\int dr \left(\nabla\phi_{e}\right)^{2} \right. \nonumber \\
    &\left. -\frac{1}{8\pi\lambda_{1}}\int dr \left(\left(\nabla\phi_{1}\right)^{2} + \kappa^{2}\phi_{1}^{2}\right)\right. \nonumber \\
    &\left. -\frac{1}{8\pi\lambda_{2}}\int dr \left(\left(\nabla\phi_{2}\right)^{2} + \kappa^{2}\phi_{2}^{2}\right)\right) \label{eq:gandcanonical2-s} \\ 
    &\times \exp\left(\frac{1}{\nu}\int dr \ln{\left(\lambda_{s}\left(\textbf{e}_{1}+\textbf{e}_{2}\right)\right)}\right) \nonumber,
\end{align}
where $\textbf{e}_{1} = \exp\left(-i\phi_{e}-i\phi_{1}\cos\theta+i\phi_{2}\sin\theta\right)$ and $\textbf{e}_{2} = \exp\left(+i\phi_{e}-i\phi_{1}\sin\theta-i\phi_{2}\cos\theta\right)$. The free energy is,
\begin{equation}
\begin{split}
    \beta\mathcal{F} = &\int dr\left[-\frac{1}{8\pi l_{B}}\left(\nabla\psi_{e}\right)^{2} -\frac{1}{8\pi\lambda_{1}}\left\{\left(\nabla \psi_{1}\right)^{2}+\kappa^{2}\psi_{1}^{2}\right\}\right. \\
    &\left.-\frac{1}{8\pi\lambda_{2}}\left\{\left(\nabla \psi_{2}\right)^{2}+\kappa^{2}\psi_{2}^{2}\right\}\right] + \eta\left(\tilde{\textbf{e}}_{1}, \tilde{\textbf{e}}_{2}\right),
\end{split}
\end{equation}
where 
\begin{equation*}
    \eta\left(\tilde{\textbf{e}}_{1}, \tilde{\textbf{e}}_{2}\right) = -\frac{1}{\nu}\int dr \ln\left\{\lambda_{s}\left(\tilde{\textbf{e}}_{1} + \tilde{\textbf{e}}_{2}\right)\right\},
\end{equation*}
and $\tilde{\textbf{e}}_{1} = \exp\left(-\psi_{e}-\psi_{1}\cos\theta+\psi_{2}\sin\theta\right)$ and $\tilde{\textbf{e}}_{2} = \exp\left(+\psi_{e}-\psi_{1}\sin\theta-\psi_{2}\cos\theta\right)$.
The functional integral is approximated by the saddle-point field, 
\begin{equation}
\begin{split}
    &\frac{1}{4\pi l_{B}}\nabla^{2}\psi_{e} = \frac{1}{\nu} \frac{\tilde{\textbf{e}}_{2}-\tilde{\textbf{e}}_{1}}{\tilde{\textbf{e}}_{1}+\tilde{\textbf{e}}_{2}} \\ \\
    &\frac{1}{4\pi \lambda_{1}}\left(\nabla^{2}-\kappa^{2}\right)\psi_{1} = \frac{1}{\nu} \frac{-\cos\theta\tilde{\textbf{e}}_{1} - \sin\theta\tilde{\textbf{e}}_{2}}{\tilde{\textbf{e}}_{1}+\tilde{\textbf{e}}_{2}} \\ \\
    &\frac{1}{4\pi \lambda_{2}}\left(\nabla^{2}-\kappa^{2}\right)\psi_{2} = \frac{1}{\nu} \frac{\sin\theta\tilde{\textbf{e}}_{1} - \cos\theta\tilde{\textbf{e}}_{2}}{\tilde{\textbf{e}}_{1}+\tilde{\textbf{e}}_{2}}.
\end{split}\label{eq:mf-s}
\end{equation}

\subsection{Molecular Dynamics Simulation}
Based on Ye's coarse-grained model, which successfully induced the divergence of differential capacitance~\cite{benjamine}, Yukawa interaction is added to compare the results with the theory. 
Cations and anions carry charges at their center. As the reference work claimed, image charge interactions are included to study the divergence of differential capacitance. In addition to the long-range Coulomb interaction, particles interact each other through combined short-ranged pair potentials. All input parameters and output quantities are scaled by fundamental quantities. The main dimensionless physical quantities are provided in section A of Supporting Information.
Simulations are performed using the LAMMPS in the canonical ensemble~\cite{lammps}. The system consists of $1000$ real particles and $1000$ image particles, with a number density of $\rho=0.8$, and the temperature is set to $T=1$ using Langevin thermostat. The simulation box is created with the size of $2L \times L \times L$ where the value of $L$ is determined to satisfy the desired number density $\rho=N/V$. Furthermore, wall potentials are applied at $x=0$ and $x=L$ to represent two smooth planar electrodes. The relative permittivity, $\epsilon_{r}$, and Bjerrum length, $l_{B}$, are set to typical values used for RTIL systems~\cite{Kornyshev, bsk, wakai}, \textit{i.e.}, $\epsilon_{r}=12$ and $l_{B}=10$. Moreover, the elementary charges have reduced magnitudes of $q=\sqrt{\epsilon_{r} l_{B} T} \simeq 11$.

Energy minimization and equilibration are performed for each simulation over $10^{5}$ timesteps with a step size $\tau=0.001$. Another $3\times10^{6} \tau$ steps are spent for data production. As Ye \textit{et al.} claimed simulation converges to the equilibrium state regardless of the initial configurations~\cite{benjamine}.

In EDLCs simulations, it is crucial to properly treat the interactions between ions and electrodes. An image charge method is an effective way to embrace the dielectric discontinuity between the electrolyte and the surrounding medium. When the relative permittivity of the bounding material is lower (higher) than that of the electrolyte, ions experience repulsion (attraction) with their images. 
The image charge method is implemented in LAMMPS by Dwelle and Willard~\cite{image_charge}. 
It is based on the method proposed by Hautman \textit{et al.}, which involves considering infinitely repeating systems of image charges when simulating a three-dimensional system with ions~\cite{rhee}.
The simulation system is divided into two subsystems: a real system on the right side ($0 < x < L_{x}$) and an image system on the left side $( -L_{x} < x < 0 )$. The image particles possess charges opposite to those of the real particles in the real system, and they are generated at positions symmetric to the real particles relative to the electrode located at $x=0$. The whole simulation box is repeated as a unit cell.

The short-range ion-ion correlation is modeled by shifted Yukawa potential. For here the potential has the form
\begin{equation}
    \beta U_{Y}(r_{ij}^{*}) =
    \begin{cases}
        \frac{A}{r_{ij}^{*}}\left(e^{-r_{ij}^{*}/l_{c}} - e^{-r_{c}^{*}/l_{c}}\right), & \,\,\,\,r_{ij}^{*} \leq r_{c}^{*} \\
        0, & \,\,\,\,r_{ij}^{*} > r_{c}^{*}
    \end{cases}
    \label{eq:yukawa-s}
\end{equation}
where $A$ represents Yukawa coefficient, which is $a$ for cation-cation, $b$ for anion-anion, and $c$ for cation-anion. $l_{c}$ is the length of the Yukawa potential.

To investigate the effect of asymmetric Yukawa potential, the correlation strength between cation-cation, $a$, is fixed to 0.75 and $b$ and $c$ are varied within certain ranges: $0.1a\leq b \leq 1.4a$ and $0.9a\leq c \leq 1.2a$. And the cutoff distance is set to $r_{c}=L_{x}/2$. Hereinafter, fractions multiplied by the anion-anion coefficient and the cation-anion coefficient will be referred to as $\alpha_{b}$ and $\alpha_{c}$, respectively. Details of rest of pair potentials are explained in the section B of Supporting Information

Subsequently, simulations with uniform electric field are performed to elucidate the effect of asymmetric Yukawa interaction on the formation of crowding structure. Electric fields with strength $-1 \leq E \leq 1$ are applied to model potential differences with strength $-118 \leq\Delta V/V_{T}\leq 118$. For here, $V_{T}=k_{B}T/e \approx 11$ is the thermal voltage in LJ units and $\alpha$, and $\alpha_{c}$ are fixed to 0.75 and 1.1, while $\alpha_{b}$ being varied.

\end{widetext}

\end{document}